\documentclass[twocolumn,aps,prc,tightenlines,floats,floatfix]{revtex4}%%

\def\etal{{\it et.\ al.\/},$\,$}

\def\bea{\begin{eqnarray}}
\def\eea{\end{eqnarray}}
\def\etal{{\it et al.\/}}

\def\etal{{\it et al.\/}}
\def\sfrac#1#2{{\textstyle \frac{#1}{#2}}}

\def\be{\begin{equation}}
\def\ee{\end{equation}}
\def\ba{\begin{eqnarray}}
\def\ea{\end{eqnarray}}

\usepackage{graphics}
\usepackage{graphicx}
\usepackage{epsf} 
\usepackage{amsmath}
\usepackage{amssymb}
\begin{document} 

\phantom{0}
\vspace{-0.2in}
\hspace{5.5in}
\parbox{1.5in}{ \leftline{JLAB-THY-09-V10}} 

\vspace{-1in}%\parbox{1.5in}{ \vspace{-9.6in}}  % moves the preprint box down

\title
{\bf 
Electric quadrupole and magnetic octupole of the $\Delta$}
\author{G. Ramalho$^{1,2}$, 
M.T. Pe\~na$^{2,3}$ and Franz Gross$^{1,4}$ 
\vspace{-0.1in}  }

\affiliation{
$^1$Thomas Jefferson National Accelerator Facility, Newport News, 
VA 23606, USA \vspace{-0.15in}}
\affiliation{
$^2$Centro de F{\'\i}sica Te\'orica e de Part{\'\i}culas, 
Av.\ Rovisco Pais, 1049-001 Lisboa, Portugal \vspace{-0.15in}}
\affiliation{
$^3$Department of Physics, Instituto Superior T\'ecnico, 
Av.\ Rovisco Pais, 1049-001 Lisboa, Portugal \vspace{-0.15in}}
\affiliation{
$^4$College of William and Mary, Williamsburg, VA 23185, USA}

\vspace{0.2in}
\date{\today}

\phantom{0}

\begin{abstract}
Using a covariant spectator constituent quark model 
we predict an electric quadrupole moment $Q_{\Delta^+}= -0.043$ 
$e$fm$^2$ and a
magnetic octupole moment ${\cal O}_{\Delta^+}= -0.0035$ $e$fm$^3$ 
for the $\Delta^+$ excited state of the nucleon.  
%\maketitle
\end{abstract}

%\phantom{0}
%\vspace{7.0in}
%\vspace{-6in}
\vspace*{0.9in}  % sets how far the title is below the preprint box
\maketitle

Although it was the first nucleon resonance to be discovered, 
the properties of the $\Delta$ are almost completely unknown.  
Only the $\Delta^{++}$ and $\Delta^{+}$ magnetic moments
have been  measured, and these measurements have large error bars
\cite{Yao06,Kotulla02,Castro01}.
Most of the information we have about the $\Delta$ comes from 
indirect information,
such as the study of the 
$\gamma N \to \Delta$ transition \cite{Blanpied01}.

The dominant $\Delta$ elastic form factors are 
the electric charge $G_{E0}$ and magnetic dipole $G_{M1}$.
The subleading form factors are 
the electric quadrupole ($G_{E2}$) and
magnetic octupole ($G_{M3}$).
Those form factors measure 
the deviation of the charge 
and magnetic dipole distribution 
from a symmetric form \cite{Buchmann08}.
At $Q^2=0$ the form factors 
define the magnetic dipole $\mu_\Delta= G_{M1}(0) \sfrac{e}{2M_\Delta}$,
the electric quadrupole $Q_\Delta=G_{E2}(0) \frac{e}{M_\Delta^2}$ 
and the magnetic octupole 
${\cal O}_\Delta= G_{M3}(0) \sfrac{e}{2M_\Delta^3}$ moments, where $e$ is the electric charge and $M_\Delta$ the $\Delta$ mass.

Until recently, there were essentially only theoretical 
predictions for $\mu_\Delta$  
(see Ref.~\cite{DeltaFF} for details) 
and $Q_\Delta$ 
\cite{Isgur82,Giannini90,Leonard90,Krivoruchenko91,Butler94,Buchmann97,Buchmann02,Arndt03}. 
The exception was the pioneering work in lattice QCD 
\cite{Leinweber92}, where all the form factors 
were estimated for low $Q^2$, although the statistics for $G_{E2}$ and $G_{M3}$ were very poor.

Recent lattice QCD calculations of all four form factors over 
a limited $Q^2$ range have revived interest in the 
$\Delta$ moments, especially the interesting quadrupole 
and octupole moments \cite{Alexandrou07,Alexandrou09}.  
%Unfortunately these 
These 
results are 
%however 
obtained only for unphysical 
pion masses in the range of 350-700 MeV so 
some extrapolation to the physical 
pion mass is required \cite{Cloet03,Pascalutsa05b}.
Still, in the absence of direct experimental information,
lattice QCD provides the best reference for 
theoretical calculations.
Stimulated by these new lattice results 
the covariant spectator quark model \cite{DeltaFF} 
and  chiral Quark-Soliton model ($\chi$QSM) 
\cite{Ledwig08} have been used to estimate 
the $\Delta$ form factors.
Simultaneously, a lattice technique 
based on the background-field method \cite{Lee05} 
has been  used to estimate the $\mu_\Delta$ 
with great precision \cite{Aubin}.  
The octupole moment ${\cal O}_\Delta$ has also been evaluated by 
Buchmann  \cite{Buchmann08} using a deformed pion cloud model,
and QCD sum rules (QCDSR) have been used to estimate both  
$Q_\Delta$ and ${\cal O}_\Delta$ \cite{Azizi08}.

The size of the moments $Q_\Delta$ and ${\cal O}_\Delta$ tells us if the $\Delta$ is deformed, and in which direction.
The nucleon, as a spin 1/2 particle, can
have no electric quadrupole moment \cite{Nucleon} 
[although the possibility remains, as pointed 
out by Buchmann and Henley \cite{Buchmann00}, 
that it might be a collective state with 
an intrinsic quadrupole moment].
%, but this would also suggest the existence of a rotational band of excited states with large electromagnetic transitions].
While the measurement of the quadrupole form factors for the  
$\gamma N \to \Delta$ transition gives some information about the deformation 
of the $\Delta$ \cite{Pascalutsa07}, 
it is very important to obtain 
an independent estimate
% of the  $\Delta$ deformation 
\cite{Alexandrou09,AlexandrouDeform}.
Motivated by these considerations, the 
Nicosia-MIT and the  Adelaide
groups are presently working on an evaluation of $G_{M3}$ using lattice QCD 
\cite{Alexandrou09,Zanotti08}.
Also  Ledwig and collaborators  are 
working in the same subject \cite{Ledwig08} using the 
$\chi$QSM.

In this Letter we use the covariant spectator formalism \cite{Gross} 
to evaluate $Q_\Delta$ and ${\cal O}_\Delta$.
Following previous work \cite{NDeltaD,LatticeD}, 
we  describe the $\Delta$ as a quark-diquark system
composed of a S-state with an admixture of two D states
\be
%\vspace{-.2cm}
\Psi_\Delta(P,k)=
N \left[
\Psi_S + a \Psi_{D3} + b \Psi_{D1} \right],
\label{eqPsiDel}
%\vspace{-.2cm}
\ee
where $a$ is the mixture coefficient of the D3 state 
($L=2$, $S=3/2$) 
and $b$ the mixture coefficient of the D1 state
($L=2$, $S=1/2$).
Each of the states are separately normalized, so that $N=1/\sqrt{1+ a^2+ b^2}$.
The S, D1 and D3 wave functions are products of spin-isospin (and, for the D states, $L=2$) operators
and an appropriate scalar wave function
$\psi_S$, $\psi_{D1}$ and $\psi_{D3}$ which 
depends only the square of the momentum $(P-k)^2$ of the off-shell quark, where $k$ is the four-momentum of the on-shell diquark \cite{NDeltaD}.

In this model 
\cite{DeltaFF,Nucleon,NDeltaD,LatticeD,FixedAxis,NDelta}
the $\Delta$ current can be written as 
\ba
J^\mu&=&
 3 \sum_\lambda \int_k \bar \Psi_\Delta(P_+,k) j_I^\mu 
\Psi_\Delta (P_-,k)\nonumber \\
&=&N^2 J^\mu_S + a N^2 J^\mu_{D3} +
 b N^2 J^\mu_{D1},
\label{eqJ2}
\ea
where $P_-$ ($P_+$) is the initial (final) $\Delta$ momentum,
and the sum is over all polarizations ($\lambda$) of the diquark, and the covariant integral $\int_k \equiv \int \sfrac{d^3k}{(2\pi)^32E_s}$ where $E_s$ is the diquark energy.
Additional terms proportional to 
$a^2N^2$, $b^2 N^2$ and $abN^2$ can be neglected
if $a$ and $b$ are small.
The quark current $j_I^\mu$ in Eq.~(\ref{eqJ2})
includes a dependence on 
the quark $u$ and $d$ charges and anomalous magnetic moments   
$\kappa_u$ and $\kappa_d$. 
%$\kappa_u=1.778$ and $\kappa_d=1.915$. 
See Refs.~\cite{DeltaFF,Nucleon} 
for details.

The current (\ref{eqJ2}) can be written 
in a standard form involving four basic 
form factors, denoted $F_i^\ast$, $i=1-4$.  
The electric and magnetic moments are linear combinations of these
\cite{DeltaFF,Pascalutsa07,Nozawa90,Weber78}, 
and at $Q^2=0$, {\it to first order in the mixing coefficients a and b\/}, 
they become  
\ba
G_{E0}(0)  &=& N^2 e_\Delta {\cal I}_S   \nonumber\\
G_{M1}(0) &=&  
N^2 f_\Delta  {\cal I}_S          
\nonumber\\
G_{E2}(0) &=&  3 (aN^2) e_\Delta {\cal I}_{D3}^\prime
\nonumber\\
G_{M3}(0) &=& 
f_\Delta N^2  \left[ a\, {\cal I}_{D3}^\prime + 
2\, b \, {\cal I}_{D1}^\prime \right],
\ea
where $f_\Delta=e_\Delta+M_\Delta \kappa_\Delta/M_N$, 
\bea
e_\Delta= \sfrac{1}{2}(1+ \bar T_3),& 
\hspace{.7cm}
&\kappa_\Delta= \sfrac{1}{2}(\kappa_+ + \kappa_- \bar T_3),
\nonumber\\
\kappa_+=2\kappa_u-\kappa_d,\;&  &\kappa_-=\sfrac23\kappa_u+\sfrac13\kappa_d,
\eea
with  $\bar T_3 = \mbox{diag}(3,1,-1,-3)$, and
\ba
%{\cal I}_S&=&  \lim_{\tau \to 0}
%\int_k \psi_S (P_+,k) \psi_S(P_-,k) \nonumber \\
{\cal I}_{D3}^\prime
&=& \lim_{\tau \to 0} \frac{1}{\tau}
\int_k b(k,q,P_+)\psi_{D3} (P_+,k) \psi_S(P_-,k) 
\nonumber \\
{\cal I}_{D1}^\prime
&=& \lim_{\tau \to 0} \frac{1}{\tau}
\int_k  b(k,q,P_+) \psi_{D1} (P_+,k) \psi_S(P_-,k),
\nonumber 
\ea
with $\tau=Q^2/(4M_\Delta^2)$ and  $b(k,q,P_+) \approx Y_{20}(\hat k)$ as 
defined in Ref.~\cite{NDeltaD}.   
The S-state wave function is normalized to unity 
(so that ${\cal I}_S=1$), and to 
{\it first order in the mixing coefficients a and b\/}, 
$N^2\to1$  so $G_{E0}(0) = e_\Delta$, giving the correct charge. 
The multipole moments E2 and M3 
are fixed by the factors ${\cal I}_{D1}^\prime$ 
and ${\cal I}_{D3}^\prime$, and are {\it zero\/} if there are no D states.
In particular $G_{E2}(0)$ is determined 
only by ${\cal I}_{D3}^\prime$,
although $G_{M3}(0)$ can depend on 
a delicate balance between   ${\cal I}_{D3}^\prime$, 
${\cal I}_{D1}^\prime$ and the  
coefficients $a$ and $b$.

%%%%%%%%%%%%%%%
\begin{table}
{\footnotesize
%\begin{center}
\begin{tabular}{l r r r r}
\hline
$G_{E2}(0)$ &  $\Delta^{++}$ &  $\Delta^+$ &  
$\Delta^0$ &   $\Delta^-$ \\
\hline
NRQM (Isgur)  \cite{Isgur82,Krivoruchenko91}  
& $-$3.82 &  $-$1.91 & 0 & 1.91 \\
NRQM 
\cite{Krivoruchenko91}  & $-$3.63    & $-$1.79   & 0 & 1.79 \\
%GP \cite{Blanpied01} & & -7.02$\pm$4.05 & & \\
Buchmann (imp) \cite{Buchmann97}   & $-$2.49 & $-$1.25 & 
  0  &  1.25 \\
Buchmann  (exc) \cite{Buchmann97}   & $-$9.28 & $-$4.64 & 
  0  &  4.64 \\
$\chi$PT \cite{Butler94}     &  $-$3.12 & $-$1.17 & 0.47  & 2.34  \\[-0.01in]
   &  $\pm$1.95 &  $\pm$0.78 & $\pm$0.20 & $\pm$1.17\\
$\chi$QSM  \cite{Ledwig08}  &         &  $-$2.15    &     &  \\
QCDSR \cite{Azizi08} &  $-$0.0452 & 
 $-$0.0226 & 0  & 0.0226 \\[-0.01in]
 & $\pm$0.0113 &  $\pm$0.0057 &  & $\pm$0.0057\\
Spectator 1 &   $-$3.87   &   $-$1.93   &  0  &   1.93 \\ 
Spectator 2 &   $-$3.36   &   $-$1.68   &  0  &   1.68 \\ 
\hline
Lattice: &   &  &  & \\
%\hline
Quenched Wilson  &  & 
\hspace{-.4cm}  $-$0.81$\pm$0.29 &   &  \\ 
Dynamical Wilson 
 & & 
\hspace{-.4cm}  $-$0.87$\pm$0.67 & &   \\ 
%Hybrid  & &   \hspace{-.5cm}
%  $-2.06$\small{$^{+0.67}_{-2.25}$} & &    \\[+0.01in]
Hybrid  & &   \hspace{-.5cm}
  $-2.06^{+1.27}_{-2.35}$ & &    \\[+0.015in]
\hline
\end{tabular}
%\end{center}
}
\vspace{-.4cm}
\caption{Summary of existing theoretical and lattice results 
for $G_{E2}(0)$. Lattice data from Ref.~\cite{Alexandrou09}.
%The result for $Q^2=0$ is an extrapolation.
Quenched Wilson has $m_\pi=411$ MeV, 
dynamical Wilson has  $m_\pi=384$ MeV, and hybrid has $m_\pi=$ 353 MeV.}
\label{tableGE2}
\end{table}
%%%%%%%%%%%%%%%

\begin{table}
{\footnotesize
%\squeezetable
\begin{center}
\begin{tabular}{l r r r r }
\hline
$G_{M3}(0)$ &  $\Delta^{++}$ &  $\Delta^+$ &  
$\Delta^0$ &   $\Delta^-$ \\
\hline
GP  \cite{Buchmann08}   & $-$11.68     &  $-$5.84 & 0 & 5.84 \\ 
QCDSR \cite{Azizi08} &  $-$0.0925  &  $-$0.0462  &  0  &  0.0462    \\[-0.04in]
  $\quad$ error & $\pm$0.0234 &  $\pm$0.0117 & &  $\pm$0.0117  \\ 
Spectator 1 &    $-$0.046  &  $-$0.023   &   0.00084  &   0.024 \\ 
Spectator 2 &   $-$3.46    &   $-$1.70    &  0.063   &  1.82 \\
\hline
\end{tabular}
\end{center}}
\vspace{-.5cm}
\caption{Summary of existing theoretical results  for $G_{M3}(0)$. 
 GP stands for general parameterization (of QCD).}
\label{tableGM3}
\end{table}
%%%%%%%%%%%%%%%

To illustrate how lattice data can be used 
to constrain models, we show results from 
two different parameterizations for the 
$\Delta$ wave functions.
The first one, denoted by 
Spectator 1 (Sp 1), is model 4 of 
Ref.~\cite{NDeltaD}. 
That model fixed the pion cloud contribution 
(using a simple parameterization) 
and adjusted the remaining valence contribution 
to fit the $\gamma N \to \Delta$ data.
The second parameterization, from Ref.~\cite{LatticeD} and  denoted
Spectator 2 (Sp 2), uses the same functional form for 
the valence part of the D-state wave functions, 
{\it but fits the valence part of the wave function 
directly to the lattice data\/} \cite{Alexandrou08b}. 
Because the pion mass used in these lattice calculations is large,
the pion cloud effects are negligible at the lattice ``point'' and 
provide a better determination of the valence 
quark contribution at that point.  
After the fit is made, the results are extrapolated 
to the physical ''point'' by replacing 
the masses of the nucleon, $\Delta$, and $\rho$ meson 
(all parameters that enter into the functional form 
of the wave functions and currents) 
to their physical masses.  
We believe that  model Sp 2 gives a more 
reliable parameterization of the $\Delta$ wave
function, but we compare it to model Sp 1 to 
show the impact of using the lattice data to constrain the fit.
In the first model (Sp 1) there is a mixture of 0.88\% 
of D3 state and 4.36\% of D1 state; the second model (Sp 2) has a mixture of 
0.72\% for both the D3 and  D1 states.

In this letter we restrict our discussion to the 
moments $Q_\Delta$ and ${\cal O}_\Delta$, which are extracted from the values of the form factors $G_{E2}$ and $G_{M3}$ at $Q^2=0$.
A more complete study will be presented in 
a future work \cite{DeltaDFF}. Our results are true predictions; once
the $\gamma N \to \Delta$ reaction has been described  no additional parameters are adjusted.
The results for $G_{E2}(0)$  are presented in 
table \ref{tableGE2} and for $G_{M3}(0)$  in table \ref{tableGM3}.  These are obtained from the integrals  ${\cal I}_{D3}^\prime=-7.00$ and ${\cal I}_{D1}^\prime= 1.59$ for Sp 1
and ${\cal I}_{D3}^\prime=-6.65$ and 
${\cal I}_{D1}^\prime= 0.24$ for Sp 2.

Before comparing our results with other models note that Sp 1 and Sp 2 give similar predictions for the quadrupole moment but very different predictions  for the octupole moment.  Clearly the octupole moment is more sensitive to the details of the model, and it is only the strong constraint imposed by the lattice data that allows us to predict that $G_{M3}^{\Delta^+}(0)\simeq-1.70$.

Both tables compare our results with 
predictions of other models.
In table \ref{tableGE2} 
we include the 
classic nonrelativistic quark model (NRQM) 
from Isgur \etal\,\cite{Isgur82},
where the tensor color hyperfine interaction 
requires a mixture of D-state quarks with 
S-state quarks. 
This description considers only 
the valence degrees of freedom, 
and the contribution for the 
electric quadrupole moment is determined 
by both the mixture coefficients and 
a confinement parameter  \cite{Giannini90,Buchmann97}. 
In these models the 
contribution for the electric quadrupole can 
be estimated in impulse approximation
\cite{Krivoruchenko91,Richard84} 
from
\be
Q_\Delta^{(imp)} = \sfrac{2}{5}e_\Delta r_n^2,
\ee
where $r_n^2$ is the neutron 
squared radius in fm$^2$.  Using a recent value of 
$r_n^2= -0.116$ fm$^2$, we obtain
$Q_{\Delta^+} \simeq -0.0464$ fm$^2$, or 
$G_{E2}^{\Delta^+}(0) \simeq -1.81$ in close agreement with the values from Ref.~\cite{Krivoruchenko91} quoted in the table.
(For a review of the earlier results, see Ref.~\cite{Krivoruchenko91}.)
Similar results are obtained by Buchmann \etal\, 
\cite{Buchmann97} using a constituent quark model 
with a D-state admixture 
\cite{Giannini90,Richard84}
with a slightly  different confinement parameterization 
and an impulse approximation to the  one-body current.

In the same work \cite{Buchmann97}, 
an estimate of 
the nonvalence contributions, based on a two-body  
exchange current 
representative of the nonvalence 
degrees of freedom, is obtained.
These nonvalence contributions are the dominant ones, 
and assuming no D-state admixture, can be estimated from
\be
Q_\Delta^{(exc)} = e_\Delta r_n^2.
\label{eqQexc}
\ee 
Although developed in the constituent quark 
formalism this relation is parameter independent 
\cite{Buchmann97}.
The expression (\ref{eqQexc}) has also been derived in the large $N_c$ limit 
\cite{Buchmann02}.
Later, the expression (\ref{eqQexc}) was 
improved using a general parameterization 
(GP) of QCD \cite{Blanpied01,Buchmann02b,Dillon99},
with the inclusion of higher order terms,
and used to extract 
$G_{E2}^{\Delta^+}(0)=-7.02\pm4.05$ from the 
$\gamma N \to \Delta$ electric quadrupole data
\cite{Blanpied01}.  
All of these results seem to suggest that 
the contribution of the pion cloud to the quadrupole moment 
could be quite large. On the other hand, calculations based on 
$\chi$PT \cite{Butler94},
and recent results derived in a  
$\chi$QSM \cite{Ledwig08}
all of which include the pion cloud, 
suggest that the pion cloud 
effect might be smaller than estimates based on Eq.~(\ref{eqQexc}).
From this we conclude that model calculations of the 
size of the pion cloud contribution 
to the quadrupole moment are inconclusive.

% FIGURE
\begin{figure}[t]
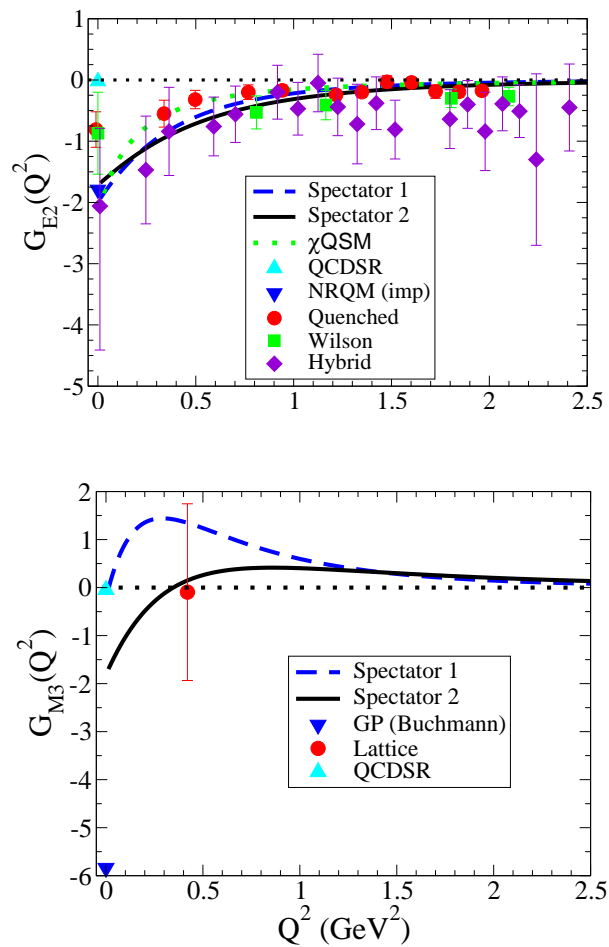

\centerline{
\mbox{
\includegraphics[width=3.1in]{GE2d}} }
%\vspace{0.75cm}
\vspace{0.9cm}
\centerline{
\mbox{
\includegraphics[width=3.1in]{GM3a}}}
\vspace{-.2cm}
\caption{\footnotesize{
$G_{E2}$ and $G_{M3}$ form factors for $\Delta^+$.
The $G_{E2}$ lattice data is from  Ref.~\cite{Alexandrou09} and 
$G_{M3}$ lattice data is from Ref.~\cite{Alexandrou07}.
The lattice points for $Q^2=0$ are result of an extrapolation \cite{Alexandrou09}.}}
%\vspace{-1cm}
\label{figGE2GM3}
\end{figure}

Finally, the tables show the lattice QCD simulations  
\cite{Alexandrou09} based on three different approaches: 
a quenched calculation using a Wilson action with 
$u$ and $d$ quarks, 
a dynamical calculation using a 
Wilson action including $u$ and $d$ sea quarks,
and a hybrid action which also includes strange sea quarks. 
The lattice data 
is however limited by the significant 
error bars that  prevent an accurate 
extrapolation to $Q^2=0$ 
(assuming a dipole or an exponential dependence on $Q^2$)
\cite{Alexandrou09} and by  
heavy pion masses (which require an extrapolation in $m_\pi$).
Even so, the size of the hybrid 
calculation may be an indicator that the 
%contrarily to the  leading form factors $G_{E0}$ and $G_{M1}$ \cite{DeltaFF}
meson cloud contribution to 
$G_{E2}$ is not negligible, 
although not comparable with (\ref{eqQexc}).
%In this case we are not considering the 
%physical extrapolation. 
Quark models can be important for extrapolating the lattice data to 
$Q^2=0$ and to the physical pion mass. In any case, 
the predictions of our model should be 
compared to other calculations of the valence quark 
contributions to these moments.

The $Q^2$ dependence of the $\Delta^+$ form factors $G_{E2}$ and $G_{M3}$ 
are shown in figure \ref{figGE2GM3}.
%The effect of the pion cloud can be inferred from Eq.~(\ref{eqQexc}).
Our results are completely consistent with the 
$Q^2$ dependence of the lattice calculations 
\cite{Alexandrou07,Alexandrou09}.   Future lattice QCD simulations would be 
important for a more precise 
constraint on ${\cal O}_{\Delta^+}$.

In conclusion, using our best model (Sp 2)
we predict
\be
%\vspace{-.4cm}
Q_{\Delta^+} = -0.043 \; e \mbox{fm}^2 \hspace{.7cm}
{\cal O}_{\Delta^+}= -0.0035 \;e \mbox{fm}^3. 
%\vspace{-.1cm}
\ee
This estimate for ${\cal O}_{\Delta^+}$ lies between the negligible predictions of QCD sum rules 
and the high estimate of Buchmann \cite{Buchmann08} 
based on a pion cloud model and the GP formalism
\cite{Buchmann08,Buchmann02,Dillon99}.
As we have previously emphasized, the small result for 
${\cal O}_{\Delta^+}$ obtained from Sp 1 shows 
the importance of using the lattice data 
to constrain the model; without this constraint 
the uncertainty in our prediction of ${\cal O}_{\Delta^+}$ 
would be much larger.

Using the ``minimal electromagnetic current'' 
defined in the historical literature,
these results imply an  
oblate  form 
for both charge and magnetic distributions of the $\Delta^+$.
%Acording with Ref.~\cite{Alexandrou09} 
%this results should be compared with the 
%``natural values'' 
%associated with a pointlike spin 3/2 particle 
%($G_{E2}(0)=-3$ and $G_{M3}(0)=-1$), leading to
%$Q_{\Delta^+}=-0.077$ $e$fm$^2$ and 
%${\cal O}_{\Delta^+}=-0.0021$ $e$fm$^3$.
However, the electromagnetic coupling recently described by Alexandrou
\etal\,\cite{Alexandrou09}, predicts that even 
a point-like $\Delta$ will have ``natural'' moments 
of $Q_{\Delta^+} = -0.077$ ($G_{E2}^{\Delta^+}(0)=-3$) 
and ${\cal O}_{\Delta^+}= -0.0021$ 
($G_{M3}^{\Delta^+}(0)=-1$) 
leading to a different interpretation of our results.

%\vspace{0.3cm}
\vspace{0.3cm}
\noindent
{\bf Acknowledgments}

This work was partially support by Jefferson Science Associates, 
LLC under U.S. DOE Contract No. DE-AC05-06OR23177.
G.~R.\ was supported by the Portuguese Funda\c{c}\~ao para 
a Ci\^encia e Tecnologia (FCT) under the grant  
SFRH/BPD/26886/2006. 
This work has been supported in part by the European Union
(HadronPhysics2 project ``Study of strongly interacting matter'').

\end{document}